\documentclass{aa}  

\usepackage{graphicx}
\usepackage{txfonts}
\begin{document} 

   \title{Suzaku observations of the low surface brightness cluster A76}


   \author{N. Ota \inst{1}\fnmsep\thanks{\email{naomi@cc.nara-wu.ac.jp}}
          \and
          Y. Fujino \inst{1}
          \and 
          Y. Ibaraki \inst{1}
          \and 
          H. B\"{o}hringer\inst{2}         
          \and
          G. Chon\inst{2}
           }

   \institute{Department of Physics, Nara Women's University, Kitauoyanishimachi, Nara, Nara 630-8501, Japan
         \and
         Max-Planck-Institut f\"{u}r  extraterrestrische Physik, Giessenbachstra\ss e, 85748 Garching, Germany   
             }

   \date{Received --; accepted --}

 
  \abstract
  {We present results of {\it Suzaku} observations of a nearby galaxy
    cluster \object{A76} at $z=0.0395$. This cluster is characterized
    by extremely low X-ray surface brightness and is hereafter
    referred to as the LSB cluster. }
  {To understand the nature and thermodynamic evolution of the LSB
    cluster by studying the physical properties of the hot
    intracluster medium in \object{A76}.}
  {We conducted two-pointed {\it Suzaku} observations of \object{A76}
    and examined the global gas properties of the cluster by XIS
    spectral analysis. We also performed deprojection analysis of
    annular spectra and derived radial profiles of gas temperature,
    density and entropy out to approximately 850~kpc ($\sim 0.6
    r_{200}$) and 560~kpc ($\sim 0.4 r_{200}$) in A76 East and A76
    West, respectively. }
  {The measured global temperature and metal abundance are
    approximately 3.3~keV and 0.24~solar, respectively. From the
    deprojection analysis, the entropy profile is found to be flat
    with respect to radius.  The entropy within the central region
    ($r<0.2r_{200}$) is exceptionally high ($\sim 400~{\rm
      keV\,cm^{2}}$). This phenomenon is not readily explained by
    either gravitational heating or preheating.  The X-ray morphology
    is clumped and irregular, and the electron density is extremely
    low ($10^{-4}-10^{-3}~{\rm cm^{-3}}$) for the observed high
    temperature, suggesting that \object{A76} is in the early phase of
    cluster formation and the gas compression due to gravitational
    potential confinement is lagging behind the gas heating.}
   {}

   \keywords{galaxies: clusters: individual: Abell~76 -- galaxies:
     intracluster medium -- X-rays: galaxies: clusters -- cosmology:
     observations}

   \maketitle
%

\section{Introduction}\label{sec:intro}
With the exception of rare dramatic cluster mergers, galaxy clusters
feature closely self-similar X-ray surface brightness profiles, with a
steep decrease with radius
\citep[e.g.,][]{2004A&A...428..757O,2008A&A...487..431C}.  A small
fraction of clusters (5--10\%) exhibit very low surface brightness
distribution and highly diffuse X-ray emission, as found e.g., in
cluster samples detected in the {\it ROSAT} All-Sky Survey (RASS)
\citep{2001A&A...369..826B}. Among representative sub-samples of
galaxy clusters from the RASS-based REFLEX survey of 33 objects
\citep[the REXCESS project;][]{ 2007A&A...469..363B}, three such LSB
clusters were identified.

A possible interpretation of the above-mentioned characteristic is
that the objects are dynamically very young. In most clusters, the
cluster core region presumably forms first, and subsequently, a large
cluster mass accretes. By contrast, these objects may develop from
rarer overdensities, which more closely resemble overdense homogeneous
spheres. Such objects would show a nearly homogeneous collapse with
high degree of local clumping.

This scenario can be validated from the entropy distribution of the
intracluster medium (ICM) in these cluster systems
\citep[e.g.,][]{2005RvMP...77..207V}.  The three LSB clusters in the
REXCESS sample are distinguished by high ($\gtrsim 200~{\rm
  keV\,cm^2}$) central entropy. This is puzzling, since the above
scenario requires an intermediate stage with very high central entropy
which must be lowered again for normal cluster
formation. Alternatively, we note that the temperature of the LSB
clusters is higher than that of normal clusters at a given gas
density. When a cluster collapses, the deepening of its potential well
is manifested by an increase in temperature (which approximately
reflects the potential depth). The gas should be simultaneously
compressed by potential confinement. The very high entropy suggests
that, by some unknown mechanism, the gas compression is lagging behind
the heating of the gas. Alternatively, unexpected heating processes
may have boosted the entropy and inflated the ICM in these systems.

To probe the nature of the LSB clusters and further test the above
scenarios, we here focus on the extremely LSB galaxy cluster
\object{A76}, which exhibits irregular morphology.  Indeed, the
surface brightness of \object{A76} is the lowest among the {\it ROSAT}
clusters studied in \cite{1999A&A...348..711N}.  The main immediate
objective of these studies is to assess the temperature and entropy
structure of the clusters using {\it Suzaku} satellite data
\citep{2007PASJ...59S...1M}. The aim is to better understand the
formation process and history of these objects.

In this study, the cosmological model is adopted with the matter
density $\Omega_{M}=0.27$, the cosmological constant
$\Omega_{\Lambda}=0.73$, and the Hubble constant $H_0=70~{\rm
  km\,s^{-1}\,Mpc^{-1}}$.  At the cluster redshift ($z=0.0395$),
1\arcmin corresponds to 47~kpc. 
Unless otherwise noted, specified errors indicate the 90\% confidence
intervals.

\section{Observation and data reduction}\label{sec:obs}
\begin{table*}
\caption{Log of {\it Suzaku} observations of \object{A76}. }\label{tab1}
\centering
\begin{tabular}{llllll} \hline\hline
Target & Obs ID & Date  &\multicolumn{2}{c}{Coordinates$^{\mathrm{a}}$} & Exposure$^{\mathrm{b}}$ \\ 
 & & & RA & Dec &   [s] \\ \hline
A76 East  & 804087010 & 2009 Dec 18      & 00:40:32.0  & 06:50:14.6 & 23640 \\
A76 West & 804088010 & 2009 Dec 18,19 & 00:39:36.7  & 06:50:16.8 & 15130 \\\hline  
\end{tabular}
\begin{list}{}{}
\item[$^{\mathrm{a}}$] Pointing coordinates in J2000.
\item[$^{\mathrm{b}}$] Net exposure time after data filtering.
\end{list}
\end{table*}

  \begin{figure*}[htb]
   \centering
\rotatebox{0}{\scalebox{0.40}{\includegraphics{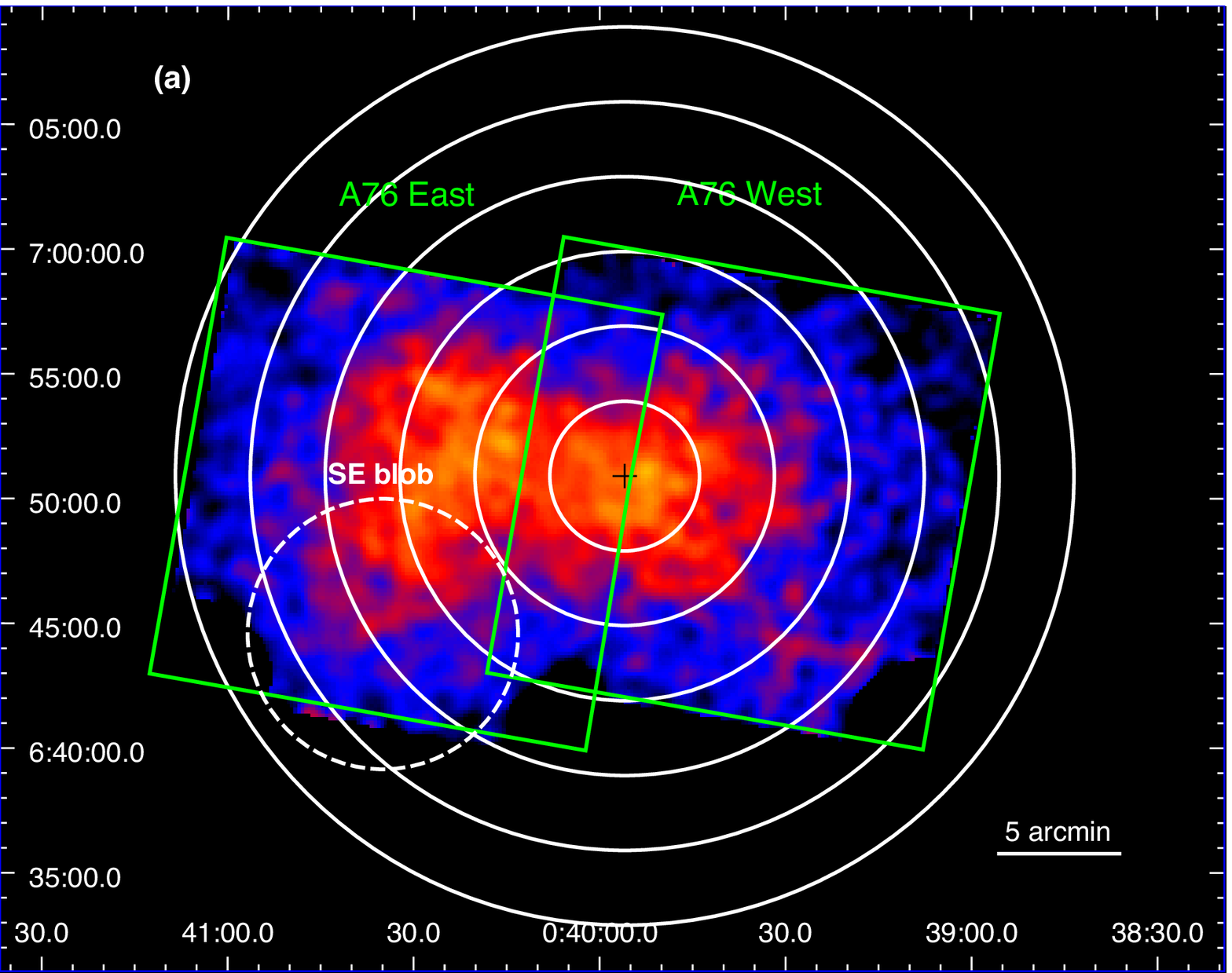}}}
\rotatebox{0}{\scalebox{0.385}{\includegraphics{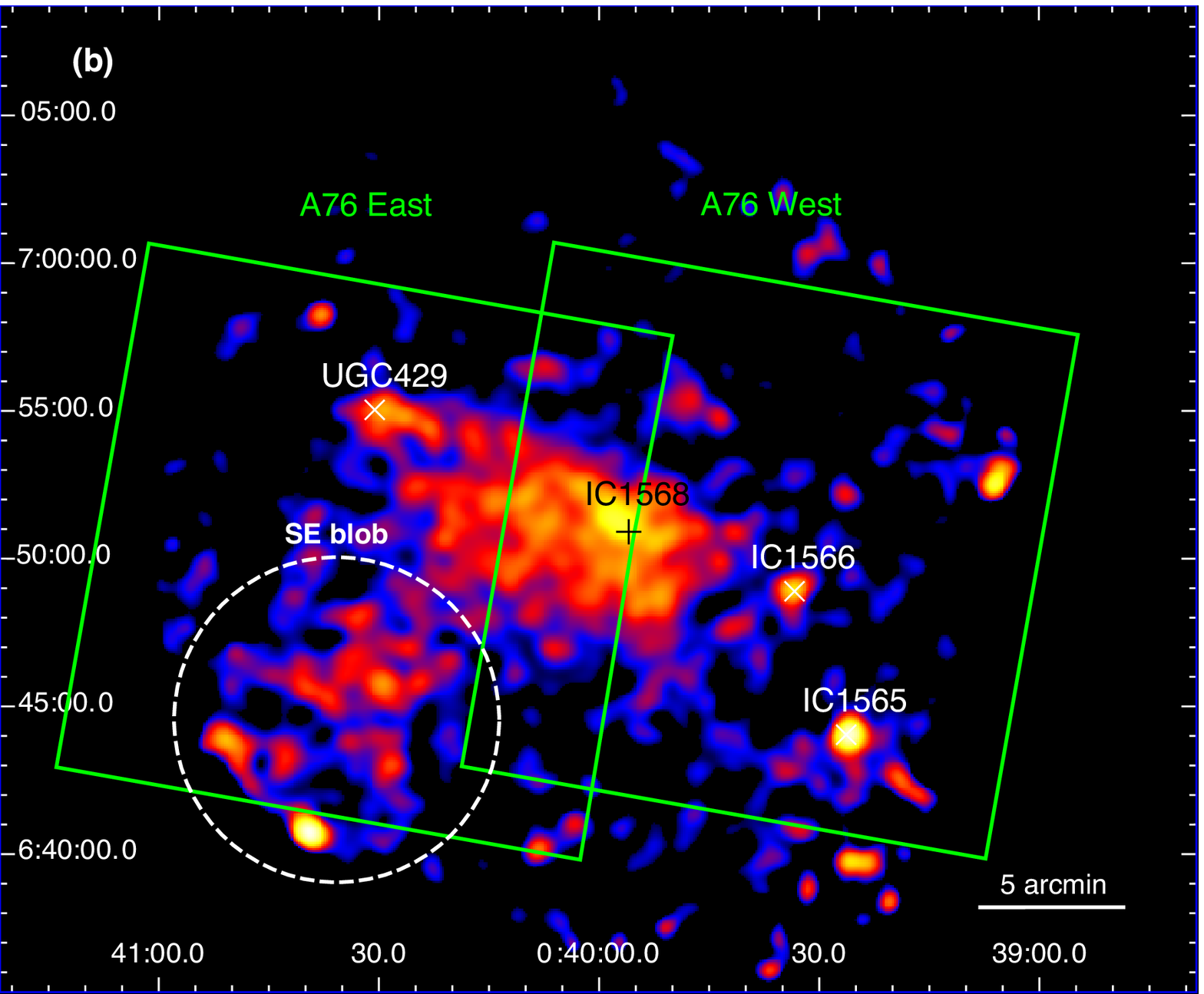}}}
\caption{(a) {\it Suzaku} XIS-1 mosaic image of \object{A76} in the
  0.5--5~keV band. The image is corrected for exposure map and
  vignetting effect and is smoothed by a Gaussian function with
  $\sigma=40\arcsec$, without background subtraction. The location of
  the X-ray peak corresponding to member galaxy IC~1568 is marked with
  the plus sign. The spectral integration regions used in
  \S\ref{subsec:global_spec} and \ref{subsec:annular_spec} lie within
  the green boxes and white circles, respectively.  The two corners of
  the CCD chip illuminated by $^{55}$Fe calibration sources are
  excluded from the image. (b) {\it XMM-Newton} PN image of
  \object{A76} in the 0.5--2~keV band with subtracted background and
  $\sigma=20\arcsec$ Gaussian smoothing.  The net exposure time is
  approximately 600~s. The positions of three detected sources
  (\S\ref{sec:obs}) and the SE blob (\S\ref{sec:discuss}) are
  indicated by crosses and dashed circle, respectively.}
   \label{fig1}%
 \end{figure*}

 Two-pointed observations of \object{A76} (in the west and east
 directions) were conducted between 2009-Dec-18 and 2009-Dec-19 in the
 {\it Suzaku} AO-4 period. The details of the observations are
 summarized in table~\ref{tab1}. The XIS instruments consist of four
 X-ray sensitive CCD cameras: three front-illuminated (XIS-0, -2, -3)
 and one back-illuminated (XIS-1) \citep{2007PASJ...59S..23K}. The
 XIS-0, -1, and -3 CCD cameras were operated in normal mode with space
 charge injection enabled \citep{2008arXiv0810.0873U}.

 As shown in Fig.~\ref{fig1}a, diffuse X-ray emission from the cluster
 is clearly detected by {\it Suzaku}/XIS. The X-ray peak position
 coincides (within the attitude determination limits of {\it Suzaku})
 with the optical coordinates of one of the member galaxies, IC~1568,
(00:39:56.0, +06:50:54.9)
 in the J2000 coordinates \citep{2008PASJ...60S..35U}.  
   Figure~\ref{fig1}b shows the {\it XMM-Newton}/PN image of the
   cluster.  The net exposure time after removing the periods of high
   background rates is about 600~s.  As is also evident from the
 image, the X-ray morphology is irregular, being elongated along the
 west- and south-east directions.

 Event files were created by pipeline processing (version 2.4). Data
 were analyzed using using HEAsoft (version 6.12) and CALDB (version
 2012-09-02 for XIS and version 2011-06-30 for the X-ray telescopes
 \citep[XRT; ][]{2007PASJ...59S...9S}).  The XIS data were filtered
 according to the following criteria: the Earth elevation angle
 $>10^{\circ}$, day-Earth elevation angle $>20^{\circ}$, and satellite
 outside the South Atlantic Anomaly.

 The cluster spectra were extracted from 1) the XIS full field of
 views and 2) annular regions centered on \object{IC~1568}.  The
 global gas properties in both pointing regions were measured from the
 former spectrum, while the latter was used to investigate the radial
 distributions.  In both cases, the two corners of the CCD chip
   covered by $^{55}$Fe calibration sources as well as circular
 regions with radius $1'$ centered on three point sources detected in
 the {\it XMM-Newton} image (\object{IC~1565}, \object{IC~1566}, and
 \object{UGC~429}) were excluded from the integration regions. The
 non-X-ray background was subtracted using {\tt xisnxbgen}
 \citep{2008PASJ...60S..11T} while other background components, i.e.,
 the cosmic X-ray background (CXB) and the Galactic emissions arising from the local hot bubble
 (LHB) and the Milky Way halo (MWH) were determined by using the same
 method described in \cite{2011PASJ...63S.979S}. In brief, the
 blank-sky data obtained from 93~ks Lockman Hole observations (Obs ID
 104002010) were modeled by the formula ``apec$_{\rm LHB}$ + wabs *
 (apec$_{\rm MWH}$ + power-law$_{\rm CXB}$)''.  The derived parameters
 are listed in Table~\ref{tab2}. The systematic error due to the
 positional dependence of the background is estimated to be 10\% by
 comparing the Lockman hole spectra with the blank-sky spectra around
 ASAS~J002511+1217.2 (OBSID 403039010), which is $\sim 6^{\circ}.6$
 offset from \object{A76}.

\begin{table*}
  \caption{Spectral fits of the Lockman hole data for the CXB and Galactic components}\label{tab2}
\centering
\begin{tabular}{lllllll} \hline\hline
$\Gamma$ & Norm  & $kT_{\rm MWH}$ & $Norm_{\rm MWH}$ & $kT_{\rm LHB}$ & $Norm_{\rm LHB}$ & $\chi^2$/d.o.f \\
                     &             & [keV] &    & [keV] &  &     \\ \hline
$1.42^{+0.04}_{-0.04}$ &  $1.05^{+0.04}_{-0.04}\times10^{-3}$ &  $0.27^{+0.05}_{-0.05}$  
& $2.0^{+0.11}_{-0.06}\times10^{-4}$ &  0.1(fix) &  $2.63^{+0.06}_{-0.07} \times 10^{-3}$  & 820/704 \\ \hline
\end{tabular}
\begin{list}{}{}
\item[$^{\mathrm{a}}$] The units are ${\rm
    photons\,keV^{-1}\,cm^{-2}\,s^{-1}}$ at 1~keV.
\item[$^{\mathrm{b}}$] Normalization of the APEC model, $Norm = \int
  n_e n_H dV/(4\pi (1+z)^2 D_A^2)~[10^{-14}{\rm cm^{-5}}]$. $D_A$ is
  the angular diameter distance to the source. An $r=20'$ uniform sky is assumed.
\end{list}
\end{table*}

The energy response files were generated by using {\tt xisrmfgen}. To
account for the vignetting effect in the XRTs, and a decrease in the
low-energy efficiency due to contaminants on the optical blocking
filter of the XIS, the auxiliary response files were calculated using
{\tt xissimarfgen} \citep{2007PASJ...59S.113I}. Here the {\it
  XMM-Newton} image of \object{A76} was used as the input surface
brightness.

\section{Analysis and results}\label{sec:analysis}
\subsection{Global spectra}\label{subsec:global_spec}
  \begin{figure*}[htb]
   \centering
\rotatebox{0}{\scalebox{0.33}{\includegraphics{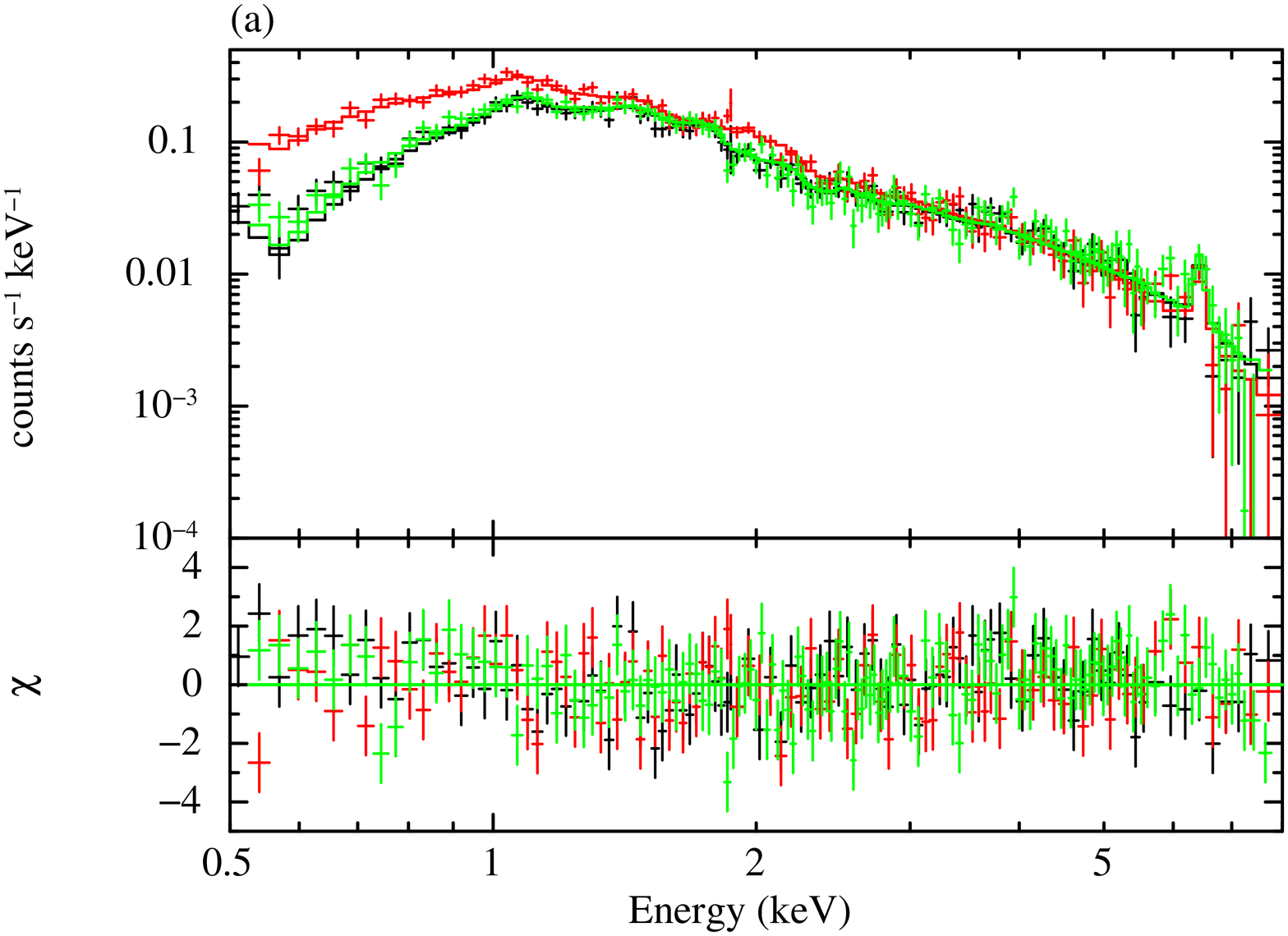}}}
\rotatebox{0}{\scalebox{0.33}{\includegraphics{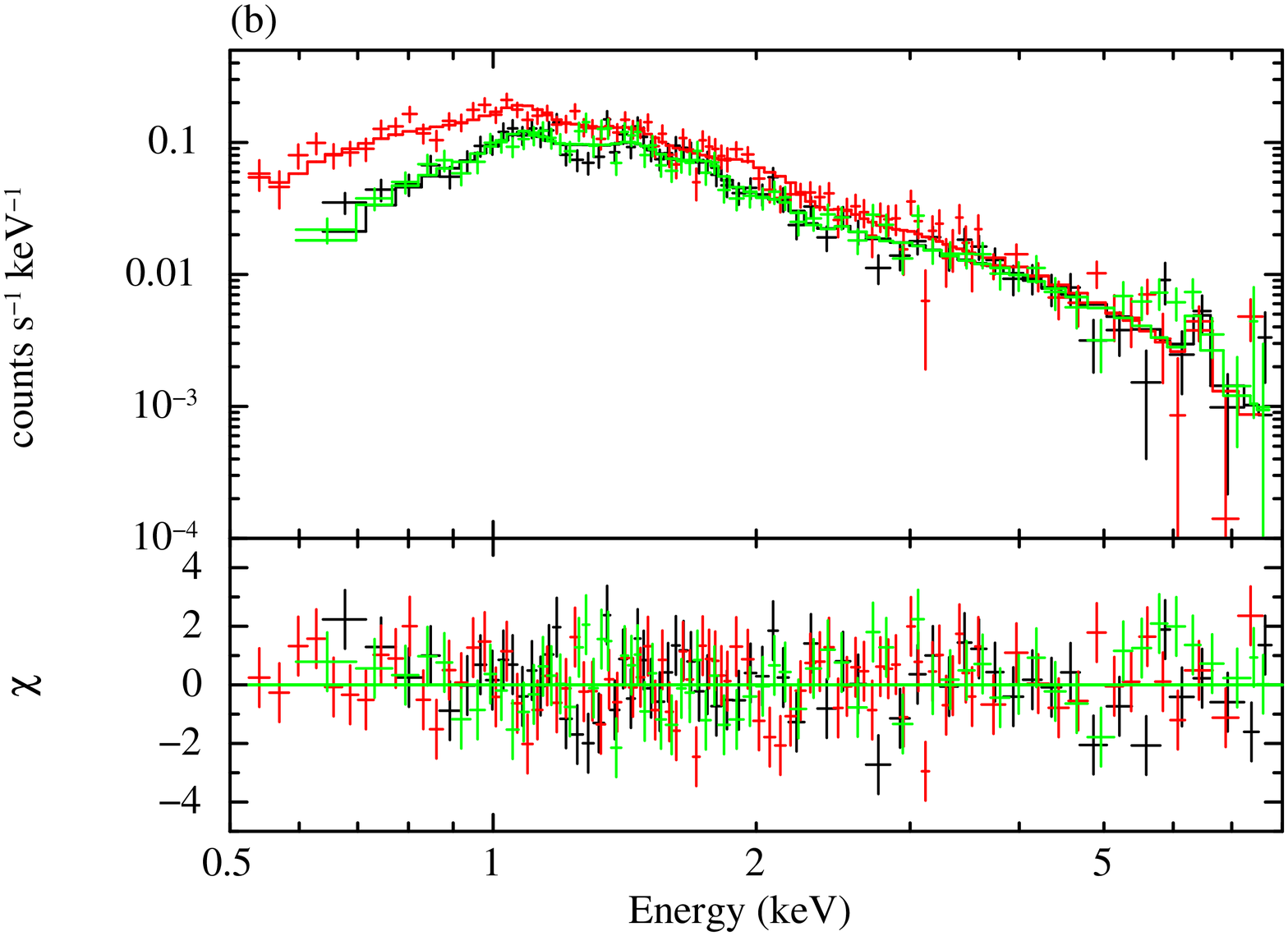}}}
\caption{Observed XIS spectra of (a)\object{A76} East and (b)
  West. The spectra collected by XIS-0 (black), 1 (red), and 3 (green)
  are shown separately. The solid lines in the upper panel show the
  best-fitting APEC model simultaneously fitted to the three sensors
  convolved with the telescope and detector response functions. Bottom
  panel shows the residuals of the fit (in number of standard
  deviations). }
   \label{fig2}%
 \end{figure*}

\begin{table*}
\caption{APEC model parameters for spectra in \object{A76} East and West}\label{tab3}
\centering
\begin{tabular}{lllll} \hline\hline
Region & $kT$~[keV] & $Z$~[solar] & $Norm$ & $\chi^2$/d.o.f \\ \hline
A76 East  & $3.39^{+0.15}_{-0.10}$ & $0.24^{+0.04}_{-0.04}$ & $1.08^{+0.03}_{-0.03}\times10^{-2}$ & 384/353 \\
A76 West  & $3.10^{+0.17}_{-0.17}$ & $0.23^{+0.08}_{-0.07}$ & $1.29^{+0.08}_{-0.07}\times10^{-2}$ & 259/219 \\ \hline
\end{tabular}
\end{table*}

To measure the average temperature of \object{A76} East and West, we
first analyzed the XIS spectra extracted from two $18'\times18'$
square regions (the solid boxes in Figure~\ref{fig1}).  The observed
0.5--8~keV spectra of three sensors (XIS-0, XIS-1, and XIS-3) were
simultaneously fitted to the APEC thermal plasma model
\citep{2001ApJ...556L..91S}. The redshift and Galactic hydrogen column
density were fixed at $z=0.0395$ and $N_{\rm H}=3.4\times 10^{20}~{\rm
  cm^{-2}}$ \citep[LAB survey;][]{2005A&A...440..775K}. For metal
abundance, the tables in \cite{1989GeCoA..53..197A} were used.  The
$\chi^2$ fitting was performed using {\tt XSPEC} version 12.7.  The
XIS spectra and model parameters are given in Figure~\ref{fig2} and
Table~\ref{tab3}, respectively.  In both regions, the temperature and
metal abundance are $kT\sim 3$~keV and $Z\sim 0.24$~solar with the
fractional errors of 4--5\% and 17--35\%, respectively.

A simultaneous APEC-model fit to the A76 East and West spectra yields
a global temperature $kT$ of $3.3\pm 0.1$~keV
($\chi^2/$d.o.f.=662/569). From the relationship given in Table~2 of
\cite{2005A&A...441..893A} and the mean cluster temperature of
$kT=3.3$~keV, the viral radius is derived as $r_{200} = 1.32$~Mpc.

\subsection{Annular spectra}\label{subsec:annular_spec}
To derive the radial temperature profile, we extracted spectra from
$3'$-wide concentric rings; $0\arcmin-3\arcmin$, $3\arcmin-6\arcmin$,
$6\arcmin-9\arcmin$, $9\arcmin-12\arcmin$, $12\arcmin-15\arcmin$, and
$15\arcmin-18\arcmin$, as shown in Figure~\ref{fig1}.  Because the
X-ray image is asymmetric, the data from \object{A76} East and West
were separately analyzed.  Note that the 6th ring
  ($15\arcmin-18\arcmin$) is beyond the range of A76 West and is used
  only in A76 East. The XIS-0,1,3 spectra of each ring were then
simultaneously fitted to the APEC model in the same manner as that
mentioned in \S\ref{subsec:global_spec}. The best-fit parameters are
listed in Table~\ref{tab4}.

The spectra of the innermost ring ($0'<r<3'$) were derived from the
\object{A76} West data alone, since the \object{A76} East data cover
only half of the area.  In the spectra of the outer rings ($15'<r<18'$
in \object{A76} East and $12'<r<15'$ in \object{A76} West), the metal
abundance was fixed at its mean value (0.24~solar) because of large
statistical uncertainty.

By summing the flux from all rings within $r<18'$, the
absorption-corrected flux and luminosity within $r<0.6r_{200}$ in the
0.5--8~keV are estimated to be $1.8\times10^{-11}~{\rm
  erg\,s^{-1}\,cm^{-2}}$ and $6.7\times10^{43}~{\rm erg\,s^{-1}}$,
respectively. The bolometric luminosity of the cluster is
$8.2\times10^{43}~{\rm erg\,s^{-1}}$ ($r<0.6r_{200}$), approximately
1/3 of the luminosity expected from the observed temperature and
luminosity-temperature relationships of nearby clusters
\citep[e.g.,][]{2009A&A...498..361P}.

\begin{table*}
\caption{APEC model parameters for the annular regions in A76 East and West}\label{tab4}
\centering
\begin{tabular}{lllllllll} \hline\hline
             & \multicolumn{4}{c}{A76 East} & \multicolumn{4}{c}{A76 West} \\ \cline{2-5} \cline{6-9}
Region & $kT$~[keV] & $Z$~[solar] & $Norm$ & $\chi^2$/d.o.f. & $kT$~[keV] & $Z$~[solar] & $Norm$ & $\chi^2$/d.o.f. \\ \hline
$0'<r<3'$ & -- & -- & -- & -- 
                  & $3.56_{-0.27}^{+0.30}$  & $0.28_{-0.13}^{+0.15}$  & $2.13^{+0.17}_{-0.16}\times10^{-3}$   &  71/78\\
$3'<r<6'$ & $3.32_{-0.20}^{+0.25}$  & $0.21_{-0.09}^{+0.10}$  & $3.55_{-0.23}^{+0.24}\times10^{-3}$  & 96/120 
                  & $3.55_{-0.34}^{+0.37}$  & $0.28_{-0.15}^{+0.17}$  & $4.08^{+0.33}_{-0.31}\times10^{-3}$  & 80/70 \\ 
$6'<r<9'$ & $3.57_{-0.25}^{+0.25}$  & $0.18_{-0.08}^{+0.09}$ 	  & $3.04_{-0.17}^{+0.18}\times10^{-3}$ & 158/155 
                  & $2.64_{-0.27}^{+0.36}$  & $0.26_{-0.13}^{+0.19}$   & $6.88_{-0.09}^{+0.10}\times10^{-3}$   & 60/64\\ 
$9'<r<12'$& $3.52^{-0.31}_{+0.30}$  & $0.19_{-0.10}^{+0.11}$  & $3.12^{+0.21}_{-0.24}\times10^{-3}$  & 140/127 
                   & $2.60_{-0.44}^{+0.56}$   & $0.33_{-0.22}^{+0.39}$  & $5.34_{-1.09}^{+1.28}\times10^{-3}$  & 31/36\\
$12' < r< 15'$ & $3.66_{-0.41}^{+0.45}$   & $0.31_{-0.18}^{+0.22}$ & $3.95^{-0.32}_{+0.34}\times10^{-3}$ &90/82 
                  & $7.1_{-3.6}^{+21.9}$ & 0.24(fix)  & $1.87_{-0.42}^{+0.68}\times10^{-3}$ & 7/8\\
$15'<r<18'$ & $ 2.79_{-0.35}^{+0.57}$  & 0.24(fix)	& $1.95^{-0.21}_{+0.24}\times10^{-3}$  & 53/45 & -- & -- & --&-- \\ \hline
\end{tabular}
\end{table*}

\subsection{Deprojection analysis}
To further derive the three-dimensional structure of the gas density
and entropy in the cluster, we performed a deprojection analysis
assuming spherical symmetry of the cluster gas distribution. Since the
X-ray morphology of \object{A76} is clearly irregular, this assumption
introduces a systematic uncertainty into the analysis, which will be
examined later in this subsection.

\begin{figure*}[htb]
   \centering
\rotatebox{0}{\scalebox{0.33}{\includegraphics{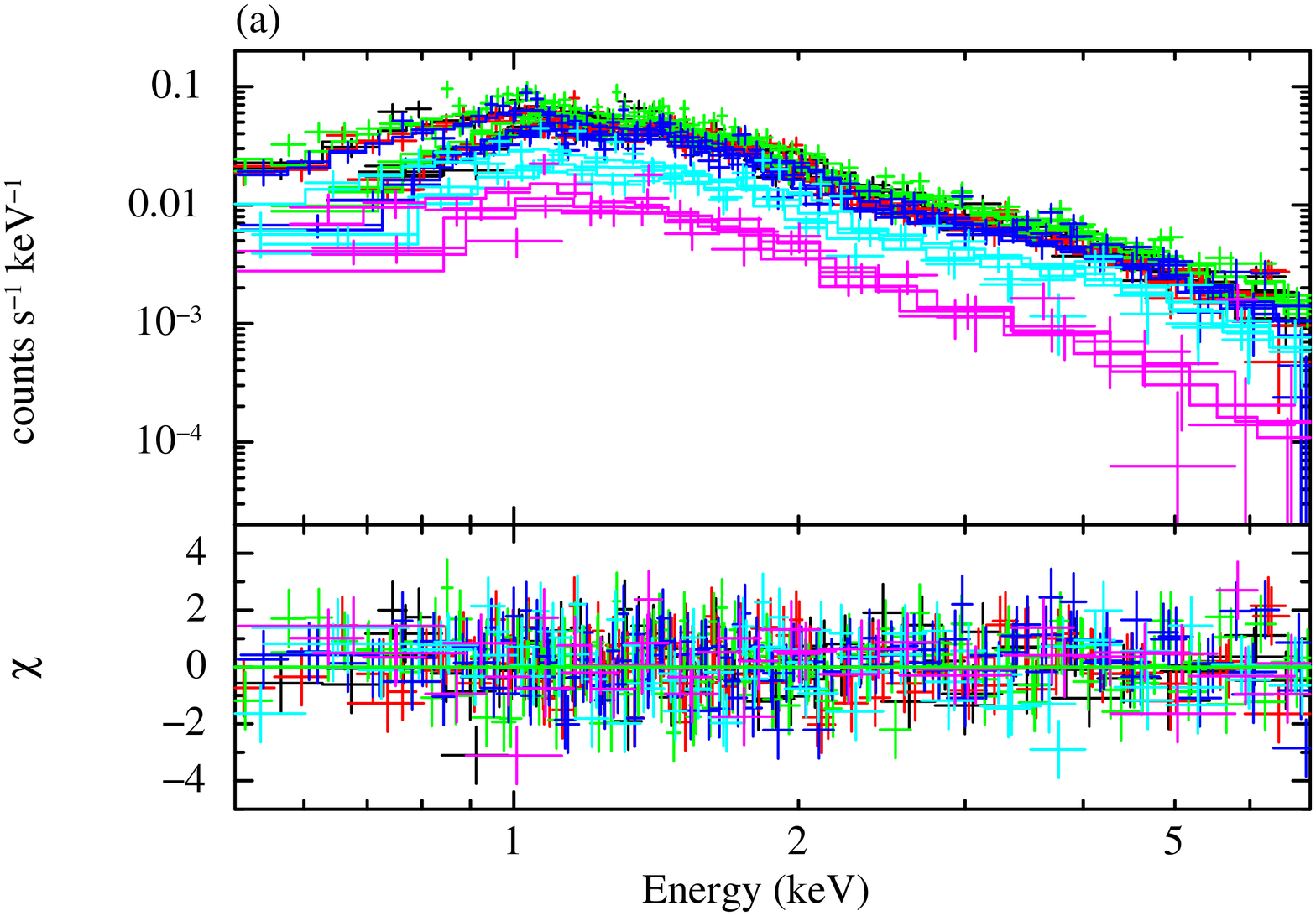}}}
\rotatebox{0}{\scalebox{0.33}{\includegraphics{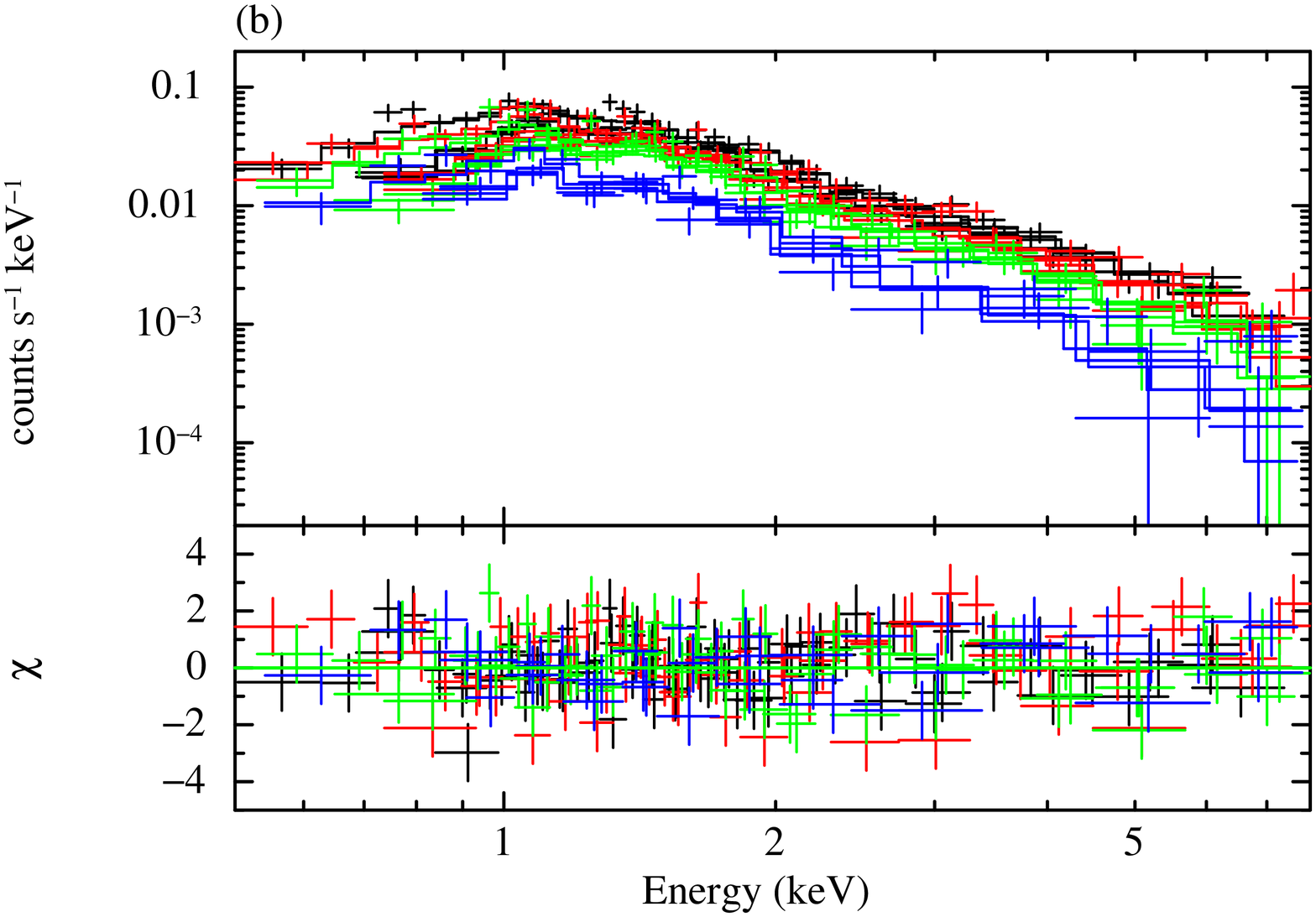}}}
\caption{Deprojection analysis of XIS spectra obtained from A76 data (a) East and
(b) West.  The crosses are the XIS-0, 1, 3 spectra accumulated from
  $0'<r<3'$ (black), $3'<r<6'$ (red), $6'<r<9'$ (green), $9'<r<12'$
  (blue), $12'<r<15'$ (cyan), $15'<r<18'$ (magenta).}
   \label{fig3}%
 \end{figure*}

The APEC model corrected for the Galactic absorption was fitted to
each radial bin, assuming that a metal abundance is radially constant.
The arithmetic deprojection operation was performed by the ``projct''
model in {\tt XSPEC}. The results of spectral fitting for \object{A76}
East and West are shown in Figure~\ref{fig3} and Table~\ref{tab5}. The
spectra of $12'<r<15'$ were excluded from the analysis of \object{A76}
West because of poor statistics, which prevented convergence of model
fit to the data.  We confirmed that inclusion/exclusion of these outer
spectra do not significantly affect the spectral parameters obtained
for the remaining regions. In addition, since the uncertainty in the
fitted temperature is large in the region $6'<r<9'$, the temperature
of that region is linked to that of the inner region ($3'<r<6'$).

\begin{figure*}[htb]
   \centering
\rotatebox{0}{\scalebox{0.33}{\includegraphics{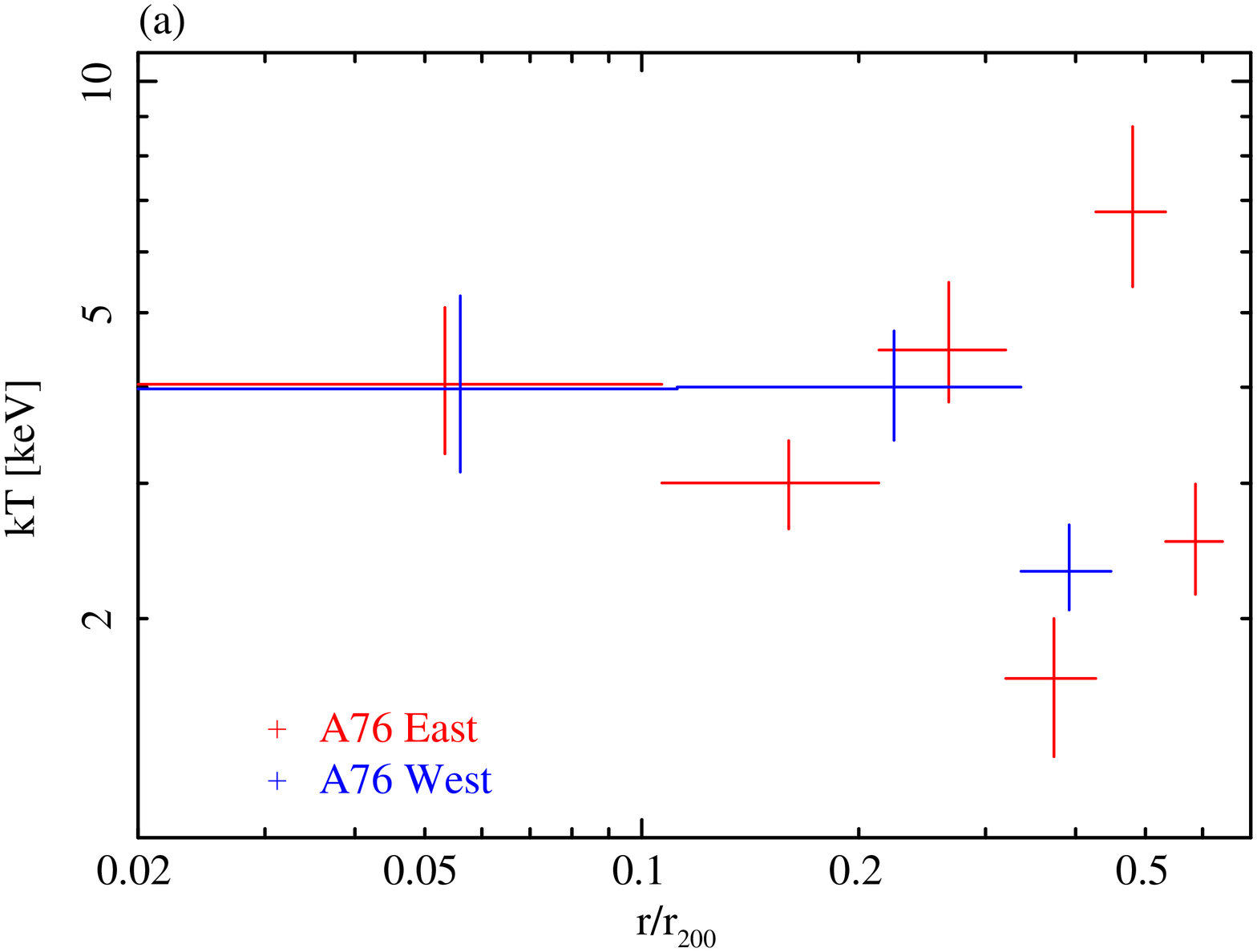}}}
\rotatebox{0}{\scalebox{0.33}{\includegraphics{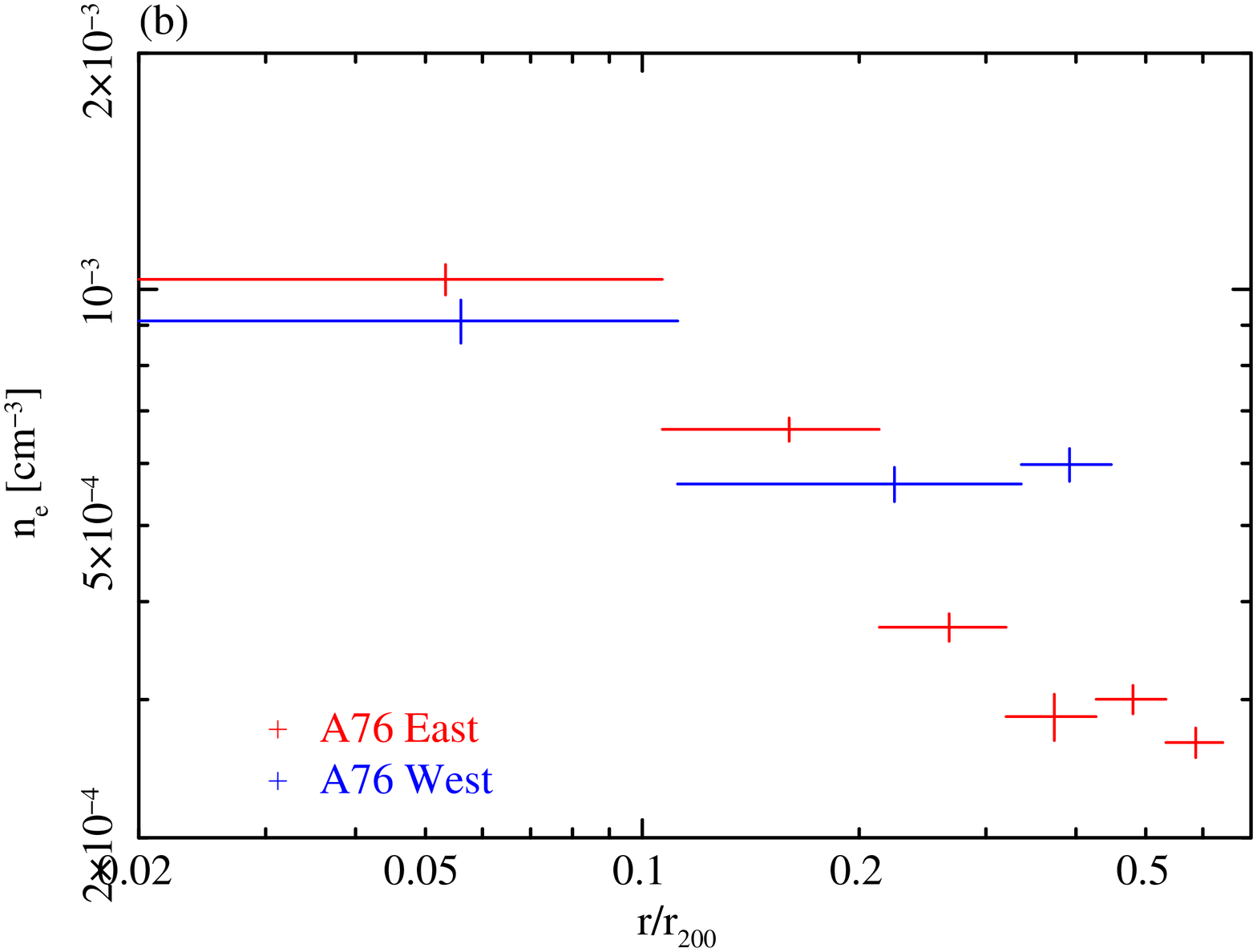}}}
\rotatebox{0}{\scalebox{0.33}{\includegraphics{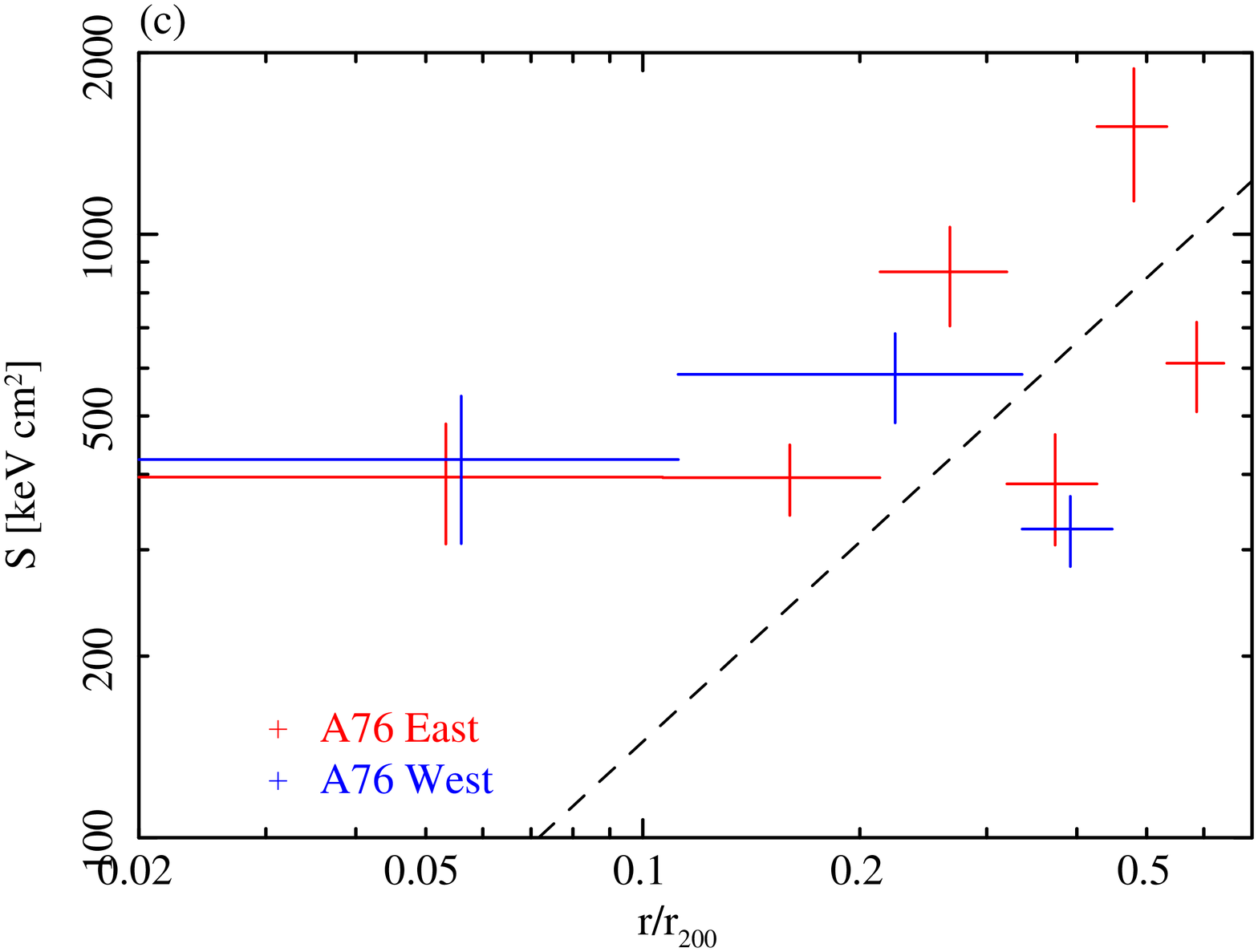}}}
\caption{(a) Temperature, (b) electron density, and (c) entropy
  profiles derived from the deprojection analysis. The results for A76
  East and West data are shown as red and blue crosses (the markers
  for A76 West are shifted horizontally by 5\% for clarity).  In panel
  (c), the dashed line represents the baseline entropy profile, $S
  \propto r^{1.1}$ \citep{2005MNRAS.364..909V}.}
   \label{fig4}%
 \end{figure*}

The derived temperature profile is shown in Figure~\ref{fig4}a. The
radius is normalized by $r_{200}$ and the electron density is
calculated from the APEC normalization factor $\int n_e n_H dV/(4\pi
(1+z)^2 D_A^2)~[10^{-14}{\rm cm^{-5}}]$, where $n_e = 1.2 n_{\rm H}$
(see Figure~\ref{fig4}b).  The measured central electron density is
approximately $10^{-3}~{\rm cm^{-3}}$, the lowest among the known
nearby clusters \citep{2008A&A...487..431C}.  The gas entropy, defined
as $S\equiv kT n_e^{-2/3}$ \citep{2005RvMP...77..207V}, is evaluated
from the above quantities within $0.6r_{200}/0.4r_{200}$ for
\object{A76} East/West (Figure~\ref{fig4}c).  The entropy profile is
found to be flat for $r<0.2r_{200}$ though large scatter occurs at
larger radii. A very high entropy of $\sim 400~{\rm keV\,cm^{2}}$ is
revealed at the cluster center.

As will be discussed in \S\ref{sec:discuss}, the entropy at the
cluster center is of particular interest. Therefore, to check the
robustness of the entropy measurement, we consider possible systematic
errors due to the following three items: (i) the uniform X-ray surface
brightness assumption, (ii) the background model and (iii) the effect
of the XRT's point spread function (PSF).  To asses the impact of (i),
we repeated the spectral analysis using arf files created by assuming
a uniform cluster surface brightness instead of the observed {\it
  XMM-Newton} image. Although the analysis increased the statistical
uncertainty in the central entropy, the result agrees with the results
listed in Table~\ref{tab6}. For (ii), the impact of background
uncertainty mentioned in \S\ref{sec:obs} was examined by intentionally
changing the background intensity by $\pm 10$\%. We find that this
does not significantly influences the results.  In assessing the
effects of (iii), the width of the annular regions were made larger
than the typical PSF of the {\it Suzaku} XRT ($\sim 2\arcmin$) to
avoid significant photon mixing between adjacent regions.  {\tt
  xissim} raytracing simulations \citep[conducted as described
in][]{2007PASJ...59..299S} confirmed that the PSF effect is as small
as $\sim 20$\% (in terms of fraction of photons or spectral
  normalization factor). Accordingly, the systematic error in the
  central entropy is estimated to be approximately 10\%, which is
  smaller than the statistical uncertainty.

\begin{table*}
\caption{APEC model parameters obtained from the deprojection analysis of A76 East and West data}\label{tab5}
\centering
\begin{tabular}{lllllllll} \hline\hline
             & \multicolumn{4}{c}{A76 East} & \multicolumn{4}{c}{A76 West} \\ \cline{2-5} \cline{6-9}
Region & $kT$~[keV] & $Z$~[solar] & $Norm$  & $\chi^2$/d.o.f. & $kT$~[keV] & $Z$~[solar] & $Norm$ & $\chi^2$/d.o.f. \\ \hline
$0'<r<3'$   & $4.04_{-0.76}^{+1.04}$$^{\mathrm{a}}$  & $0.15_{-0.05}^{+0.05}$$^{\mathrm{a}}$ & $8.99^{+0.08}_{-0.08}\times10^{-3}$ $^{\mathrm{a}}$ & 630/610
                    &$3.98_{-0.88}^{+1.28}$  & $0.25_{-0.08}^{+0.09}$ & $7.05_{-0.88}^{+0.90}\times10^{-4}$  & 299/255\\
$3'<r<6'$   & $3.00_{-0.39}^{+0.41}$  &  & $2.61_{-0.17}^{+0.19}\times10^{-3}$&  
                    & $4.00_{-0.59}^{+0.73}$  &    & $1.34_{-0.30}^{+0.30}\times10^{-3}$ &\\
$6'<r<9'$   & $4.37_{-0.65}^{+1.00}$  &  &$2.22_{-0.17}^{+0.19}\times10^{-3}$ &
                   & $4.00_{-0.59}^{+0.73}$$^{\mathrm{b}}$ &  &$5.70_{-0.63}^{+0.65}\times10^{-3}$  &\\
$9'<r<12'$ & $1.67_{-0.35}^{+0.33}$  &  & $2.55^{+0.03}_{-0.04}\times10^{-3}$  &
                    & $2.30_{-0.25}^{+0.34}$  &  & $1.12_{-0.10}^{+0.11}\times10^{-2}$ &\\ 
$12' <r<15'$& $6.76_{-1.36}^{+1.97}$ & & $4.66_{-0.04}^{+0.04}\times10^{-3}$  && -- & -- & -- & \\ 
$15'<r<18'$ & $2.52_{-0.37}^{+0.47}$   & & $5.40_{-0.04}^{+0.05}\times10^{-3}$  && -- & -- & -- & \\ \hline
\end{tabular}
\begin{list}{}{}
\item[$^{\mathrm{a}}$] $0'<r<3'$ spectra taken from the A76 West pointing were used in the deprojection analysis.
\item[$^{\mathrm{b}}$] Linked to the value for $3'<r<6'$.
\end{list}
\end{table*}

\begin{table*}
\caption{Gas density and entropy in A76 East and West}\label{tab6}
\centering
\begin{tabular}{lllll} \hline\hline
            & \multicolumn{2}{c}{A76 East} & \multicolumn{2}{c}{A76 West} \\ \cline{2-3} \cline{4-5}
Region      &  $n_{e0}~{\rm [cm^{-3}]}$ & $S~{\rm [keV\,cm^{2}]}$ &  $n_{e0}~{\rm [cm^{-3}]}$ & $S~{\rm [keV\,cm^{2}]}$\\ \hline
$0'<r<3'$   &  $(1.03 \pm 0.05)\times 10^{-3} $ & $ 395 \pm   89$ & $(9.11 \pm 0.58)\times 10^{-4} $ & $ 423 \pm  116$ \\    
$3'<r<6'$   &  $(6.63 \pm 0.23)\times 10^{-4} $ & $ 394 \pm   52$ & $(5.65 \pm 0.28)\times 10^{-4} $$^{\mathrm{a}}$ & $ 585\pm 98$$^{\mathrm{a}}$ \\    
$6'<r<9'$   &  $(3.71 \pm 0.15)\times 10^{-4} $ & $ 866 \pm  161$ & & \\                                                     
$9'<r<12'$  &  $(2.85 \pm 0.19)\times 10^{-4} $ & $ 385 \pm   80$ & $(5.98 \pm 0.29)\times 10^{-4} $ & $ 324 \pm   43$ \\    
$12'<r<15'$ &  $(3.00 \pm 0.13)\times 10^{-4} $ & $ 1508 \pm 373$ & -- & -- \\                                                     
$15'<r<18'$ &  $(2.65 \pm 0.12)\times 10^{-4} $ & $ 611 \pm  103$ & -- & -- \\  \hline                                                   
\end{tabular}
\begin{list}{}{}
\item[$^{\mathrm{a}}$] Values obtained from region $3'<r<9'$ in A76 West.
\end{list}
\end{table*}

\section{Discussion}\label{sec:discuss}
The temperature, gas density, and entropy profiles derived from {\it
  Suzaku} data were well constrained out to $0.6 r_{200}$ and $0.4
r_{200}$ in \object{A76} East and West directions, respectively.  In
this section, the entropy profile of \object{A76} is compared with
those of nearby clusters and theoretical predictions, and the origin
of high entropy in the LSB cluster \object{A76} is discussed.

\subsection{Comparison of A76 entropy profile with those of other clusters}
\begin{figure*}[htb]
\centering
\rotatebox{0}{\scalebox{0.33}{\includegraphics{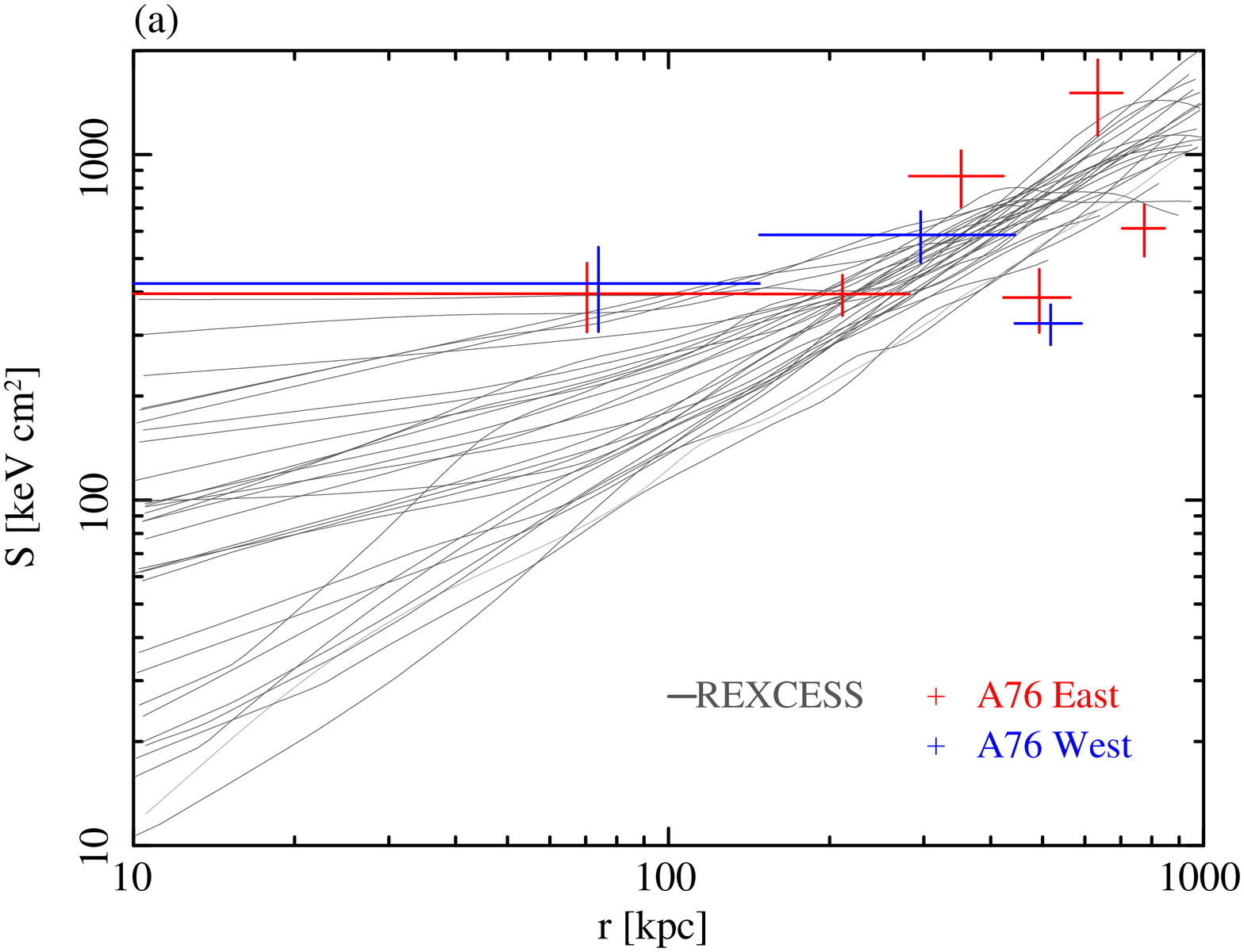}}}
\rotatebox{0}{\scalebox{0.33}{\includegraphics{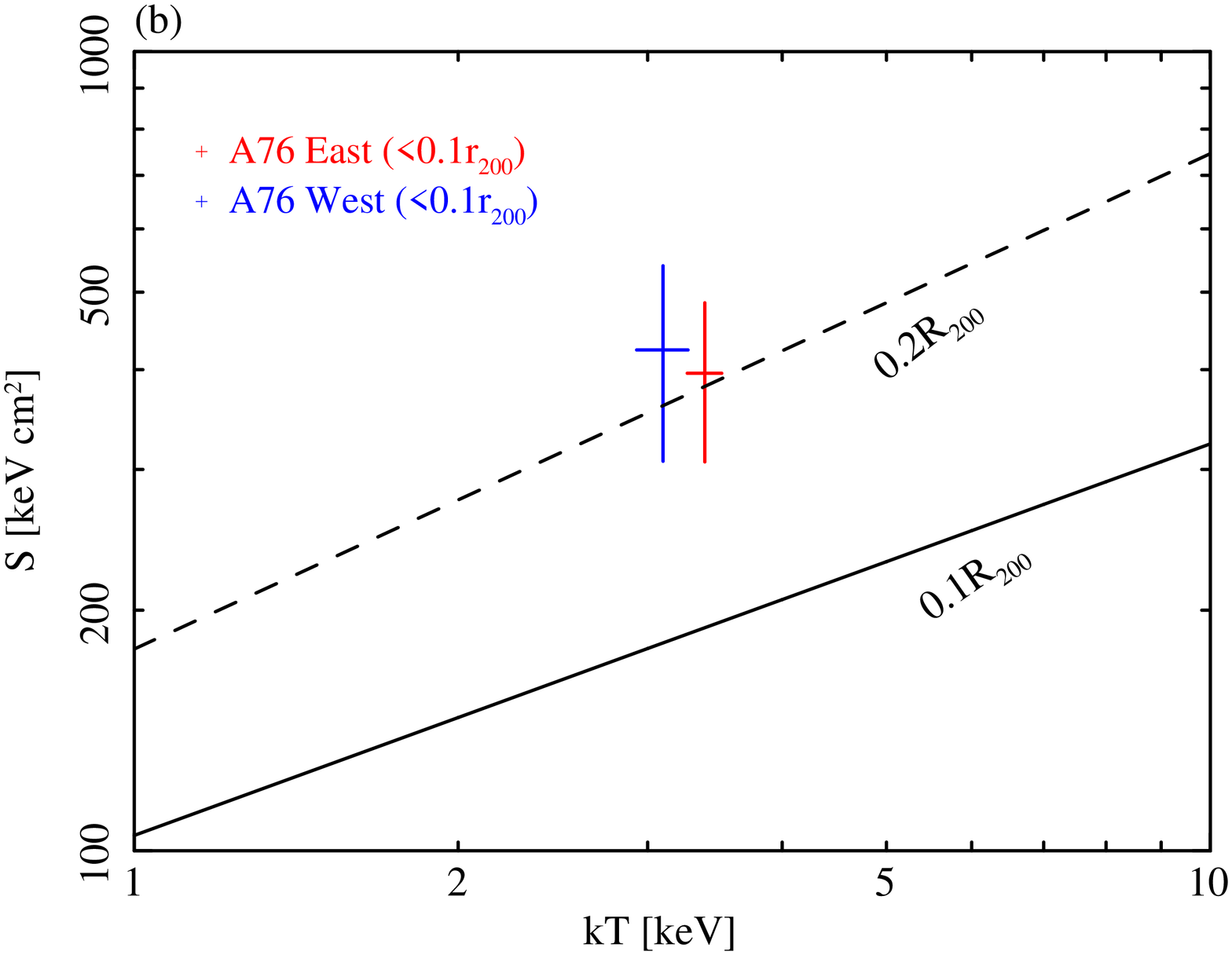}}}
\caption{(a) Comparison of radial entropy profiles between A76 and the
  REXCESS sample \citep[gray lines;][]{2010A&A...511A..85P}.  (b)
  Correlation between entropy and temperature. The red and blue
  crosses show the {\it Suzaku} results for A76 East and West data,
  respectively, while the solid and dashed lines show the $S-T$
  relationships of 10 morphologically relaxed clusters at $0.1r_{200}$
  and $0.2r_{200}$ \citep{2006A&A...446..429P}.  }
   \label{fig5}%
 \end{figure*}

\cite{2010A&A...511A..85P} derived the entropy profiles of the REXCESS
sample from {\it XMM-Newton} data.  Comparing our results with theirs
(Figure~\ref{fig5}a), we find that \object{A76} exhibits one of the
highest entropies at $r \lesssim 100$~kpc.  Comparison with other
  nearby clusters observed by {\it Suzaku} \cite[for review,
  see][]{2013SSRv..tmp...33R} yields similar results.  On the other
hand, the entropy increase at larger radii follows the overall trend
of the REXCESS sample though the \object{A76} data show significant
scatter.

For a more quantitative analysis of the central entropy,
Figure~\ref{fig5}b plots the mean gas temperature of the A76 East and
West data (Table~\ref{tab3}) and the entropy at $r<3\arcmin \sim
0.1r_{200}$ on the $S-T$ plane.  The mean $S-T$ relations at two
different radii, $0.1r_{200}$ and $0.2r_{200}$, derived from 10
morphologically relaxed clusters using {\it XMM-Newton} are also
plotted \citep{2006A&A...446..429P} .  This figure clearly shows that
the central entropy in \object{A76} is significantly higher than that
expected for the same temperature at $0.1r_{200}$.

\subsection{Comparison with the baseline profile and possibility of preheating and AGN feedback}
\cite{2005MNRAS.364..909V} conducted simulations on gravitational
heating and reported that the radial gas entropy profile in clusters
tends to follow a power-law, $S(r)= 550~{\rm keV\,cm^{2}}
(T_{200}/{\rm 1~keV})r^{1.1}$, which can be used as a baseline for
assessing the impact of non-gravitational processes in the ICM.
Figure~\ref{fig4}c compares the observed \object{A76} profile with the
baseline profile calculated for $kT_{\rm 200}=3.3$~keV. At
$0.1r_{200}$, the observed entropy ($\sim 400~{\rm keV\,cm^{2}}$) is
two times more than that expected from gravitational heating alone
($150~{\rm keV\,cm^{2}}$).

Non-gravitational heating is a plausible mechanism, particularly in
smaller systems such as groups of galaxies. According to
\cite{2003MNRAS.343..331P}, excess entropy is of the order of
$140~{\rm keV\,cm^{2}}$. On the basis of {\it Chandra} analysis of 21
groups and 19 clusters, \cite{2010RAA....10.1013W} suggested that the
entropy excess $\Delta K_0$ correlates with the K-band luminosity
$L_{K}$ of the central dominating galaxies in the systems and is
attributable to AGN feedback. Utilizing the K-band luminosity
$\log(L_K/L_{K,\odot}) =11.5$ for IC~1568 and the relationship shown
by \cite{2010RAA....10.1013W} $\Delta K_0 \propto L_K^{1.6}$, the
entropy excess is estimated as $\Delta K_0 \sim 1~{\rm
  keV\,cm^{2}}$. Therefore, the AGN feedback as well as preheating are
an unlikely cause of the high entropy generation in \object{A76}.

\subsection{Implication on the origin of high entropy in A76}
In \S\ref{sec:analysis}, the gas properties were derived mainly from
the radially averaged spectra. The X-ray image, however, reveals an
inhomogeneous gas distribution. This suggests that the cluster is at
an early stage of its formation. The lowest gas density at a
relatively high temperature of 3--4~keV indicates that gas compression
due to gravitational potential confinement is lagging behind the gas
heating. Given the low electron density of $10^{-4}-10^{-3}~{\rm
  cm^{-3}}$, the timescale of radiative cooling is significantly
longer than the Hubble time across the cluster.  At such high central
entropy level, the evolution of the gas in the deepening potential
should be chiefly adiabatic. In this scenario, the entropy would
unusually remain high even if the cluster existed in a relaxed state,
a phenomenon not observed in other relaxed clusters.  This discrepancy
may be an artifact if the gas in the central region is heavily
clumped. Low entropy clumps in the high entropy region could settle at
the center as the cluster relaxes.  As shown in the {\it XMM-Newton}
image of Figure~\ref{fig1}, the ICM indeed appears clumpy, with small
denser regions particularly common in the halos of several
galaxies. This ICM could segregate to the center and form a lower
entropy core during cluster relaxation.

A clumpy substructure is particularly apparent in the south-east
region of the cluster (hereafter, referred to as the SE blob).  From
the XIS spectra of the $r=5\arcmin.5$ circular region enclosing the SE
blob (Figure~\ref{fig1}), $kT$ is obtained as $3.2\pm0.2$~keV.  The
density and entropy are estimated as
$n_{e0}=(7.1\pm0.3)\times10^{-4}~{\rm cm^{-3}}$ and $S=397\pm32~{\rm
  keV\,cm^{2}}$, assuming that the emission comes from a uniform
sphere of radius $r=5\arcmin.5$. Both temperature and density are
significantly high within the same radial range $(0.3-0.5)r_{200}$ in
\object{A76} East (Figure~\ref{fig4}), resulting in an entropy
comparable to that obtained at the same radius from the cluster center
in other directions.  The above arguments and the spatial extent
suggest that the SE blob is probably a group-scale system that has
undergone gas heating in the cluster potential.

\section{Summary}\label{sec:summary}
We have performed X-ray spectral analysis of the LSB cluster
\object{A76}, which is characterized by extremely low X-ray surface
brightness.  We have constructed the profiles of gas temperature,
density, entropy from the {\it Suzaku} satellite data, and out to
large radii ($0.6r_{200}$ and $0.4r_{200}$ in \object{A76} East and
\object{A76} West data, respectively) for the first time. The central
gas entropy ($\sim 400~{\rm keV\,cm^{2}}$) is one of the highest among
the known nearby clusters and significantly higher than that predicted
from the mean $S-T$ relationship.  Given that the gas density is
extremely low for the observed high temperature of 3--4~keV as well as
that the gas distribution in the cluster is irregular and clumpy, we
suggest that \object{A76} is at an early stage of cluster formation
and that gas compression caused by potential confinement is lagging
behind the gas heating. Currently \object{A76} is the only LSB cluster
to be analyzed using {\it Suzaku} data. To further clarify the nature
of LSB clusters and their thermodynamic evolution, we consider that
many more galaxy clusters must be sampled and analyzed.

\begin{acknowledgements}
  We are grateful to the {\it Suzaku} team members for satellite
  operation and instrumental calibration. We would also like to thank
  G. W. Pratt for supplying the entropy profiles for the REXCESS
  sample and K. Sato for suggesting a useful means of calculating the
  PSF effect. This study is in part supported by a Grant-in-Aid by the
  Ministry of Education, Culture, Sports, Science and Technology,
  22740124 (NO).  GC acknowledges support from Deutsches Zentrum f\"ur
  Luft und Raumfahrt with the program ID 50 R 1004. GC and HB
  acknowledge support from the DFG Transregio Program TR33 and the
  Munich Excellence Cluster ``Structure and Evolution of the
  Universe''.
\end{acknowledgements}

\bibliographystyle{aa}
\bibliography{a76_aa}

\end{document}